\documentclass[final,3p,times,twocolumn]{elsarticle}

\usepackage{amssymb}

\journal{thermochimica acta}

\begin{document}

\begin{frontmatter}



\title{Thermal Analysis and Phase Diagram of the LiF-BiF$_3$ System}


\author[IPEN,IKZ]{G. H. G. Nakamura\corref{cor1}}
\cortext[cor1]{corresponding author}
\ead{gdynkm@gmail.com}
\author[IKZ]{D. Klimm}
\author[IPEN]{S. L. Baldochi}

\address[IPEN]{Instituto de Pesquisas Energ\'eticas e Nucleares, IPEN/CNEN-SP, Av.Prof. Lineu Prestes, 2242, CEP 05508-000, S\~ao Paulo, SP, Brazil}
\address[IKZ]{ Leibniz Institute for Crystal Growth, Max-Born-Stra\ss e 2, 12489 Berlin, Germany}


\begin{abstract}

Differential thermal analysis up to complete melting was performed for the complete pseudobinary system LiF--BiF$_3$. Melts with high bismuth fluoride concentration show severe evaporation, nevertheless even pure BiF$_3$ could be molten at $655^{\,\circ}$C. The system contains one intermediate compound BiLiF$_4$ which melts by peritectic decomposition under the formation of LiF at $415^{\,\circ}$C. The eutectic between BiLiF$_4$ and BiF$_3$ melts at $415^{\,\circ}$C. By thermodynamic assessment the following parameters were found for BiLiF$_4$: $\Delta H=-1514900$\,J\,mol$^{-1}$, $S=158.5$\,J\,mol$^{-1}$\,K$^{-1}$, $c_\mathrm{p}=166.173 -0.01072\cdot T$ (in J\,mol$^{-1}$\,K$^{-1}$). The melt has negative excess Gibbs free energy.



\end{abstract}


\end{frontmatter}


\section{Introduction}
\label{sec:Introduction}

Already 1987 Schultheiss et al. \cite{Schultheiss87} reported the Czochralski growth of BiLiF$_4$ single crystals up to 70\,mm length. These crystals were found to have scheelite structure, in analogy to the intermediate RELiF$_4$ compounds that were reported to exist in many rare earth fluoride--lithium fluoride systems \cite{Thoma66b}. BiLiF$_4$ was reported to melt incongruently at $369^{\,\circ}$C, but no data about the whole pseudo-binary system LiF--BiF$_3$ are available so far. However, the ternary Li-Bi-F was evaluated theoretically based on elementary electrochemical principles recently \cite{Doe09}, and it was concluded that two intermediate ternary compounds exist, namely BiLiF$_4$ and BiLiF$_6$. This is a confirmation that only for BiLiF$_4$ bismuth is trivalent \cite{Schultheiss87}.

The component LiF is well known and can be handled without special precautions, of course some care should be taken as all fluorides are more or less subject of hydrolysis under too humid conditions. It is a well known compound \cite{Baldochi01,Thoma61}, which does not undergo polymorphism upon heating, is not hygroscopic and melts congruently at $842^{\,\circ}$C.

BiF$_3$, however, is by far more sensible: It is reported that it has orthorhombic crystal structure (tysonite structure, $Pnma$ space symmetry group) \cite{Greis77,Greenwood98}. However, the melting point given there \cite{Greis77} is with $757^{\,\circ}$C $\sim100$\,K higher than the value that is measured in this study. An older study reports BiF$_3$ to be cubic \cite{Hund49} (gananite, pdf 39-345), and this is in agreement with X-ray spectra that were performed in the framework of the current study. BiF$_3$ is a stable chemical under ambient conditions, aside from minor hygroscopicity, and melts congruently at about $649^{\,\circ}$C without polymorphism \cite{Doe09,Cubicciotti68}.

The different data on BiF$_3$ may be explained by a susceptibility of contamination of the compound by oxygen and humidity, a trait common to many fluorides \cite{Greenwood98}. Degradation of samples, crucibles and even thermal sensors employed during measurements are cited by numerous authors \cite{Schultheiss87,Cubicciotti68,Darnell68,Novikova81}, hence worsening reproducibility of data. It should be mentioned that the materials used in those works -- vitrified carbon, platinum, nickel -- are known for being resistant to the chemical corrosion of fluorides, which are very aggressive at high temperature \cite{Hagenmuller85}.

Novikova {\em et al.} \cite{Novikova81} suggested that the degradation of samples containing BiF$_3$ is due to the reduction of this compound by the materials of the crucibles and sensors upon heating. Schultheiss, {\em et al.} \cite{Schultheiss87}, however, presented detailed evidence that pure BiF$_3$ is not responsible for the damage. Instead, the presence of contaminants (oxygen and water related ions) leads to partial hydrolysis of BiF$_3$ and may result in damage by the formation of Bi$_2$O$_3$ or metallic bismuth.

BiLiF$_4$ could be potentially interesting for scintillation detector applications. Natural bismuth is monoisotopic $^{209}$Bi, and due to this large atomic mass, the compound should efficiently interact with high energy photons. This paper reports on investigations of the LiF--BiF$_3$ system by thermal analysis and X-ray diffraction and is considered as preliminary step for future studies on BiLiF$_4$, particularly due to the significant crystal growth difficulties reported by Schultheiss, {\em et al.} \cite{Schultheiss87}.

\section{Experimental}
\label{sec:Experimental}

Differential thermal analysis (DTA) and differential scanning calorimetry (DSC) measurements were performed with a Netzsch STA Jupiter 449C with vacuum-dense rhodium furnace under dynamic argon atmosphere (99.999\% purity, 30\,ml/min flow rate). Graphite crucibles with platinum lids and compounds of high purity were used: BiF$_3$ 99.999\% (metals basis) from Alfa Aesar and LiF from Aldrich, initially only 99.9\% pure. The LiF, however, was refined by zone melting under Ar/HF flow \cite{Guggenheim63}. The ``type S'' thermocouples of the sample carriers were calibrated at the melting points of Zn and Au, and at the phase transition point of BaCO$_3$, respectively. Degassing of the samples and the chamber was performed under vacuum of approximately $5\times10^{-5}$\,mbar for a few hours prior to each experiment, a precautionary measure to purge the chamber and samples of oxygen and humidity, which are particularly present in BiF$_3$.

Little or no damage to the graphite components was observed throughout the many experiments, in contrast to Schultheiss {\em et al.} \cite{Schultheiss87} who reported damage for low purity compounds and/or humidity. Platinum parts, however, showed signs of corrosion after prolongued use. Samples in the entire composition range, 0--100\% BiF$_3$, were measured by DTA, in steps of 5\%; they were of approximately 50\,mg. Most of the measurements were performed with a thermal rate of 10\,K/min; further measurements with selected samples were performed with 2\,K/min in order to resolve overlapping peaks.

X-ray powder diffraction measurements of larger samples ($\sim250$\,mg) were made in a General Electric XRD 3003TT diffractometer of Bragg-Brentano geometry using Cu K$\alpha$ radiation.

\section{Results}
\label{sec:Results}

\begin{figure}[htb]
\centering
\includegraphics[width=0.48\textwidth]{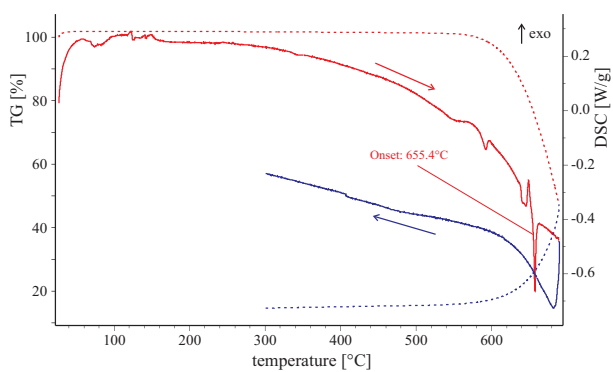}
\caption{DSC and TG curves (heating and cooling with 10\,K/min) of pure BiF$_3$.}
\label{fig:BiF3}
\end{figure}

A DSC measurement of pure Bismuth Fluoride was performed at a rate of 10\,K/min. The resulting curve is presented in Fig.~\ref{fig:BiF3}, together with the thermogravimetry curve. Significant mass loss occurred above 575$^{\,\circ}$C, which reflects the known high vapor pressure at higher temperatures which amounts to ca. 20\,mbar at this temperature \cite{Cubicciotti68}. Several minor endothermic peaks are present in the curve in that range, likely related to heat loss due to sublimation. The sharp peak around 655$^{\,\circ}$C corresponds to melting which became obvious by visual inspection of the sample after the measurement. This temperature corroborates some of the literature \cite{Cubicciotti68,FactSage6_2}, though it differs from the value given by the producer Alfa Aesar, 727$^{\,\circ}$C. It is remarkable that this value is almost identical with the $\alpha$-to-$\delta$ transition of Bi$_2$O$_3$ that has a high transformation heat of 29.8\,kJ/mol \cite{FactSage6_2,Yashima05}. It seems realistic to assume that the too high melting temperature reported by Alfa Aesar and some other references \cite{Greis77} has its origin in partial hydrolysis of BiF$_3$ to the oxide prior to, or during the measurements. The results of thermoanalytic measurements with LiF were entirely compatible with the literature.

\begin{figure}[htb]
\centering
\includegraphics[width=0.48\textwidth]{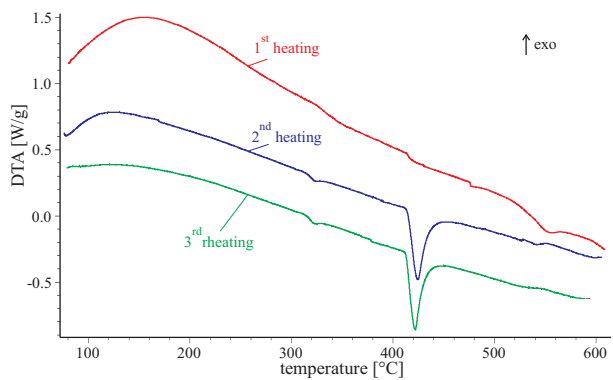}
\caption{First, second, and third heating runs (10\,K/min) of a LiF--BiF$_3$ sample with 65\% BiF$_3$.}
\label{fig:triple}
\end{figure}

Samples of intermediate composition could not be prepared by mixing them prior to the measurement e.g. in a mortar, because hydrolysis with ambient humidity or close contact with other materials would contaminate them. Instead, the components were weighed directly in the crucibles and were mixed by melting them together in a first DTA heating/cooling cycle. Fig.~\ref{fig:triple} shows such first DTA mixing run together with the subsequent heating cycles for one sample.

\begin{figure}[htb]
\centering
\includegraphics[width=0.48\textwidth]{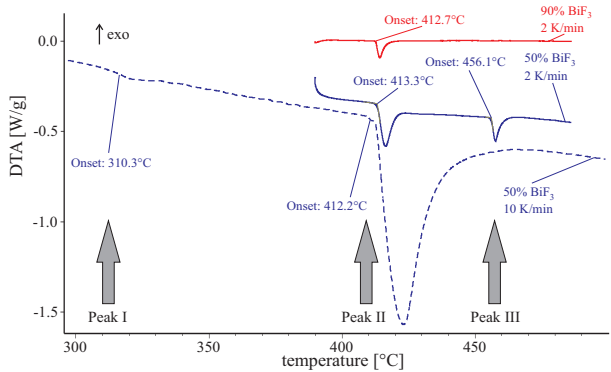}
\caption{The three invariant peaks for different compositions and heating rates.}
\label{fig:peaks}
\end{figure}

Three thermal events of invariant temperatures were observed for the different compositions within the LiF--BiF$_3$ system. An intense transition was observed at approximately 415$^{\,\circ}$C in almost all compositions (peak~II). By visual inspection of samples that were cooled after passing peak~II it became obvious that it is due to melting of a major part of the sample. Another transition (peak~III) was detected at ca. 450$^{\,\circ}$C. The separation of both peaks~II and III was only possible at slower heating rates of 2\,K/min; at 10\,K/min both peaks were overlapping. Peak~III was clearly observed only in samples with bismuth content lower than 60\%. For higher concentrations of BiF$_3$ it was either not observed, or only minor. The smallest ``peak~I'' was observed during nearly all measurement over the whole concentration range between LiF to BiF$_3$ at a constant temperature of approximately 315$^{\,\circ}$C, but did never occur for the pure components. Examples for the three invariant peaks are shown in Fig.~\ref{fig:peaks}, and the thermal events are summarized in Table~\ref{tab:peaks}. In Fig.~\ref{fig:peaks} for the 50\% BiF$_3$ sample peaks~II and III are separated for the thermal rate of 2\,K/min, but overlapping hides peak~III in the case of the higher thermal rate; it does not occur at all in the 90\% BiF$_3$ sample, however.

\begin{table*}[ht]
\begin{center}
\begin{tabular}{lll}
\hline
transition temperature              & composition range     & remark \\
\hline
$842^{\,\circ}$C                    & 0\%                   & LiF melting \\
$\sim315^{\,\circ}$C (invariant)    & $0< x< 100$\% BiF$_3$ & peak I \\ 
$\sim415^{\,\circ}$C (invariant)    & $0< x< 100$\% BiF$_3$ & peak II \\
$\sim450^{\,\circ}$C (invariant)    & $x< 60$\% BiF$_3$     & peak III \\
$655^{\,\circ}$C                    & 100\% BiF$_3$         & BiF$_3$ melting \\
\hline
\end{tabular}
\caption{Relevant transitions observed in the thermoanalytic measurements.}
\label{tab:peaks}
\end{center}
\end{table*}

\begin{figure}[htb]
\centering
\includegraphics[width=0.48\textwidth]{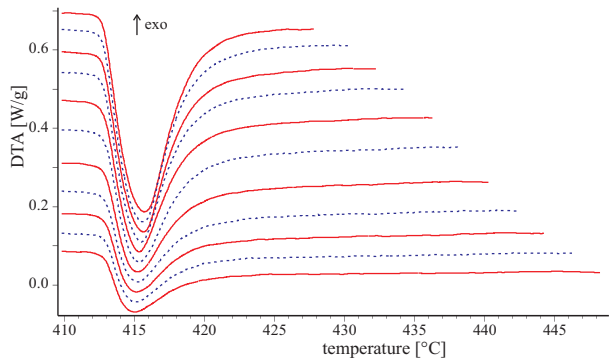}
\caption{From top to bottom: 11 subsequent heating cycles (heating/cooling runs with $\pm2$\,K/min; sample with 45\% BiF$_3$). For each subsequent cycle the maximum temperature $T_\mathrm{max}$ is increased by 2\,K. The topmost cycle has $T_\mathrm{max}=428^{\,\circ}$C, the last cycle shown has $T_\mathrm{max}=448^{\,\circ}$C.}
\label{fig:cycles}
\end{figure}

It turned out that the position for all three peaks were highly reproducible ($\sim\pm2$\,K) though all measurements, but not so the peak area. A strong tendency to smaller peaks after repeated heating/cooling cycles was evident, when some critical temperature range beyond peak~II was reached. This is demonstrated in Fig.~\ref{fig:cycles}. For this measurement a 45\% BiF$_3$ sample of 44.27\,mg was heated with 20\,K/min to a starting temperature $T_\mathrm{s}=390^{\,\circ}$C. Then a series of 45 heating/cooling cycles with rates $\pm2$\,K/min was performed up to a maximum temperature $T_\mathrm{max}=(400+2i)^{\,\circ}$C was performed where $i$ is the number of the cycle, starting from $i=0$. Fig.~\ref{fig:cycles} shows the heating runs from $i=14$ ($T_\mathrm{max}=428^{\,\circ}$C) to $i=24$ ($T_\mathrm{max}=448^{\,\circ}$C). The heating curves for all previous heating cycles $i<14$ (not shown in the figure) where almost identical; thus demonstrating that the sample was not affected by hydrolysis or evaporation during the long measurement time exceeding 40 hours in total. Instead, the following mechanism is proposed to be responsible for the shrinking peak~II area:

Bismuth fluoride has a very high density of 8.2--8.3\,g\,cm$^{-3}$. (This value from the thorough work by Hund and Fricke \cite{Hund49} seems more reliable than other sources claiming lower values around 5.3\,g\,cm$^{-3}$.) In contrast, the density of LiF is only 2.6\,g\,cm$^{-3}$. If one of the components crystallizes from the molten mixed phase, it either settles on the bottom (BiF$_3$) or floats on the surface (LiF). A part of the substance is removed from equilibrium reactions this way. This process continues with every heating/cooling cycle and can be observed visually by an inhomogeneous color of larger samples (see below).

X-ray powder diffraction patterns of two samples of 250\,mg mass with the composition of 40\% BiF$_3$ were taken. Both were prepared in the furnace of the DTA equipment by annealing at 475$^{\,\circ}$C, because under these conditions contamination with air or humidity could be ruled out. The first sample was cooled from this temperature with a rate of 10\,K/min and its diffraction pattern yielded nothing but the components of the system, LiF and BiF$_3$, despite obvious signs of melting. Segregation of the components in the sample was visible to the eye, due to differing coloration, with LiF (which is white) concentrated on top, and BiF$_3$ (grey) on the bottom of the crucible. This is surprising, as the formation of the intermediate compound BiLiF$_4$ was expected.

\begin{figure}[htb]
\centering
\includegraphics[width=0.48\textwidth]{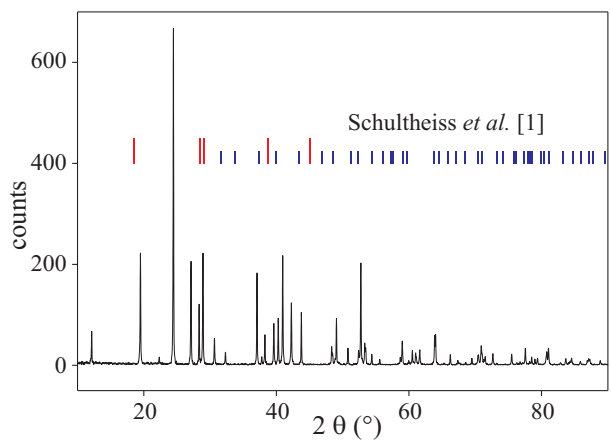}
\caption{X-ray powder diffraction pattern of a 40\% BiF$_3$ sample. The vertical bars indicate the positions that would follow from PDF entry 01-085-1491 (scheelite type BiLiF$_4$, \cite{Schultheiss87}. The longer red bars correspond to the 5 strongest scheelite peaks.}
\label{fig:X-ray}
\end{figure}

The second sample was cooled from 475$^\circ$C at a considerably slower thermal rate of 2\,K/min and yielded an unknown phase. Unexpectedly, the diffraction pattern of this phase is different from the scheelite pattern \cite{Schultheiss87} (Fig.~\ref{fig:X-ray}). One can conclude that clearly an intermediate phase made of LiF and BiF$_3$ formed that has no scheelite structure. From the current data it was impossible to relate the measured pattern to a crystal structure. Nevertheless polymorphism of the compound BiLiF$_4$ seems a realistic assumption, because e.g. the scheelites YLiF$_4$ \cite{Grzechnik02} and LuLiF$_4$ \cite{Grzechnik05b} become monoclinic at high pressure. Schultheiss \emph{et al.} \cite{Schultheiss87} performed Czochralski growth with scheelite YLiF$_4$ seeds. Such conditions might enforce the crystallization of BiLiF$_4$ in the same crystal structure type, which is not necessarily the thermodynamic stable one under the given conditions.

\section{Phase Diagram}
\label{sec:Phase Diagram}

The construction of the phase diagram will be described in comparison with the LiF--REF$_3$ (RE: rare earth element) phase diagrams that were presented e.g. by Thoma \cite{Thoma66b}. In these systems, the REF$_3$ are the highest melting compounds with $T_\mathrm{f}\approx1200-1400^{\,\circ}$C, depending on the RE element. LiF melts at $842^{\,\circ}$C (Tab.~\ref{tab:peaks}). For the elements with smallest RE$^{3+}$ radius (Lu, Yb, Tm, Er) the intermediate scheelite compound RELiF$_4$ melts congruently near $800^{\,\circ}$C. For the next larger RE (Ho, Dy, Tb, Gd, Eu) $T_\mathrm{f}$ grows monotonously for the REF$_3$, and the RELiF$_4$ undergo peritectic melting under the formation of the corresponding REF$_3$. The peritectic decomposition temperature tends to be lower for the larger RE ($799^{\,\circ}$C for Ho, only $690^{\,\circ}$C for Eu). For the next two RE's the scheelites are even more unstable and undergo a peritectoid decomposition to solid LiF and solid REF$_3$ at $570^{\,\circ}$C (SmLiF$_4$) or $362^{\,\circ}$C (PmLiF$_4$), respectively.

For LiF--BiF$_3$ the situation is reversed, because BiF$_3$ melts substantially lower than LiF. Incongruent melting of BiLiF$_4$ was already reported \cite{Schultheiss87}, but it is reasonable that this process leads to the peritectic formation of LiF instead of BiF$_3$. As the strong thermal event peak~II was clearly due to melting is is realistic to assume that this is the eutectic temperature. The somewhat weaker peak~III occurring for LiF-rich compositions (Tab.~\ref{tab:peaks}) results from the peritectic decomposition of BiLiF$_4$. Liquidus temperatures could unfortunately be detected only for a few compositions mainly in the LiF-rich part of the diagram, because the strong evaporation for BiF$_3$-rich samples did also influence the DTA signal (Fig.~\ref{fig:BiF3}).

\begin{figure}[htb]
\centering
\includegraphics[width=0.48\textwidth]{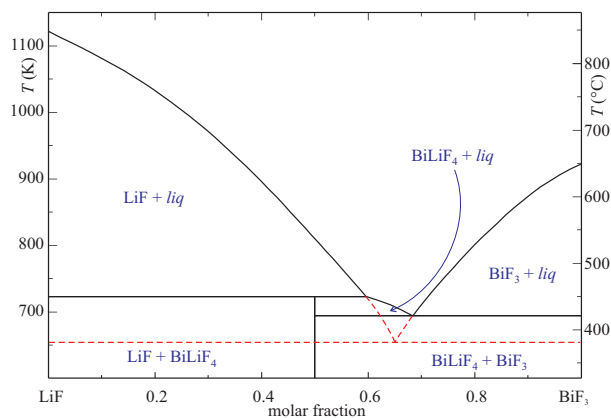}
\caption{Full lines: The equilibrium phase diagram LiF--BiF$_3$. Dashed lines: Non-equilibrium formation of a LiF/BiF$_3$ eutectic if BiLiF$_4$ formation is incomplete due to segregation.}
\label{fig:PD}
\end{figure}

The equilibrium phase diagram in Fig.~\ref{fig:PD} explains all experimental results. Solid lines result from a thermodynamic assessment of the system with the FactSage thermodynamic program and database system \cite{FactSage6_2}. Data for LiF and BiF$_3$ could be extracted directly from the ``Fact53'' database. Raw data for BiLiF$_4$ were calculated with the ``mixer'' module (Neumann-Kopp rule) and then refined to $\Delta H=-1514900$, $S=158.5$, $c_\mathrm{p}=166.173 -0.01072\cdot T$. The excess Gibbs free energy of the melt was assessed in a polynomial (Kohler/Toop) model to
\begin{eqnarray}
\Delta G_\mathrm{ex} &=& [x (1-x) (2922 - 12.77\,T - 0.005\,T^2)] \nonumber \\
                     & &+ [x (1-x)^2 (-1033 - 4.444\,T -0.01\,T^2)]
\end{eqnarray}
and is negative for all conditions shown in Fig.~\ref{fig:PD} (all data in J or J\,mol$^{-1}$\,K$^{-1}$, respectively).

The invariant point with lowest temperature (peak~I in Table~\ref{tab:peaks}) is tentatively interpreted as a metastable eutectic between LiF and BiF$_3$: The formation of BiLiF$_4$ either by the peritectic reaction or from the solid components is a diffusion controlled, and hence slow process. Moreover, segregation of LiF (top) and BiF$_3$ (bottom) resulting from large density difference allows that both components are available also if the system resides within the bottom phase fields of Fig.~\ref{fig:PD} where BiLiF$_4$ is an equilibrium phase. Consequently minor amounts of LiF and BiF$_3$ can be available inside the crucible at the same time as non-equilibrium phases and form a eutectic. The position is estimated by the intersection of the LiF and BiF$_3$ liquidus prolongation which is shown in Fig.~\ref{fig:PD} by dashed lines. Actually the temperature found there ($\sim380^{\,\circ}$C) is higher than peak~I ($\sim315^{\,\circ}$C, Tab.~\ref{tab:peaks}), but it should be taken into account that this temperature was not subject of the current (equilibrium) assessment. Besides, no liquidus data for BiF$_3$ rich melts were available, and hence the liquidus slope there relies basically on the FactSage data for its heat of fusion ($\Delta H_\mathrm{f}=21.6$\,kJ\,mol$^{-1}$) \cite{FactSage6_2}. A smaller $\Delta H_\mathrm{f}$ would shift the eutectic to lower $T$.

\section{Conclusions}
\label{sec:conclusions}

It is possible to perform DTA/TG measurements of LiF--BiF$_3$ mixture with arbitrary composition up to the melting point in lidded graphite crucibles in dry flowing argon. For pure BiF$_3$ and BiF$_3$-rich mixtures, however, the melting processes is accompanied by severe evaporation. The system contains one intermediate phase BiLiF$_4$, but the X-ray powder diffraction pattern of this phase is different from the scheelite related pattern that was measured by Schultheiss \emph{et al}. \cite{Schultheiss87} for this compound. This discrepancy could originate from polymorphism of this phase, where the actual crystal structure depends on the growth conditions (seeded growth vs. unseeded solidification in the crucible). BiLiF$_4$ melts incongruently at $450^{\,\circ}$C under formation of LiF. This temperature is considerably higher than reported by Schultheiss \cite{Schultheiss87}, but their value $(369\pm5)^{\,\circ}$C was measured with a Ni-Cr thermocouple that was corroded during the measurement.

\section*{Acknowledgments}

The authors are indebted to A. Kwasniewski for the X-ray diffraction measurements. This work was supported by CNPq (477595/2008-1; 290111/2010-2), DAAD-CAPES (po-50752632), and PROBRAL/CAPES n.~368/11.

\section*{References}



\begin{thebibliography}{10}

\bibitem{Schultheiss87}
E.~Schultheiss, A.~Scharmann, D.~Schwabe, J. Crystal Growth 80 (1987) 261--269.

\bibitem{Thoma66b}
R.~E. Thoma, Progress in the Science and Technology of the Rare Earths, Pergamon, 1966, Ch. The rare earth halides, pp. 90--122.

\bibitem{Doe09}
R.~E. Doe, K.~A. Persson, G.~Hautier, G.~Ceder, Electrochemical and Solid-State Lett. 12 (2009) A125--A128.

\bibitem{Baldochi01}
S.~L. Baldochi, S.~P. Morato, Encyclopedia of Materials: Science and Technology, Elsevier, 2001, Ch. Fluoride Bulk Crystals: Growth, pp. 3200--3204.

\bibitem{Thoma61}
R.~E. Thoma, C.~F. Weaver, H.~A. Friedman, H.~Insley, L.~A. Harris, H.~A.~Yakel Jr., J. Phys. Chem. 65 (1961) 1096--1099.

\bibitem{Greis77}
O.~Greis, M.~Martinez-Ripoll, Z. Anorg. Allg. Chem. 436 (1977) 105--112.

\bibitem{Greenwood98}
N.~N. Greenwood, A.~Earnshaw, Chemistry of the Elements, Butterworth, Leeds, 1998.

\bibitem{Hund49}
F.~Hund, R.~Fricke, Z. Anorg. Chem. 259 (1949) 198--204.

\bibitem{Cubicciotti68}
D.~Cubicciotti, J. Electrochem. Soc. 115 (1968) 1138--1143.

\bibitem{Darnell68}
A.~J. Darnell, W.~A. McCollum, J. Phys. Chem. 72 (1968) 1327--1334.

\bibitem{Novikova81}
E.~N. Novikova, P.~P. Fedorov, G.~V. Zimina, A.~Y. Zamanskaya, Y.~V. Schirokov, S.~B. Stepina, P.~I. Fedorov, V.~E. Prokopets, B.~P. Sobolev, Russ. J. Inorg. Chem. 26 (1981) 416--418.

\bibitem{Hagenmuller85}
P.~Hagenmuller, Inorganic Solid Fluorides: Chemistry and Physics, Academic Press, New York, 1985.

\bibitem{Guggenheim63}
H.~Guggenheim, J. Appl. Phys. 34 (1963) 2482--2485.

\bibitem{FactSage6_2}
{GTT Technologies, Kaiserstr. 100, 52134 Herzogenrath, Germany}, {FactSage} 6.2, \ttfamily http://www.factsage.com/ \normalfont (2010).

\bibitem{Yashima05}
M.~Yashima, D.~Ishimura, K.~Ohoyama, J. Am. Ceram. Soc. 88 (2005) 2332--2335.

\bibitem{Grzechnik02}
A.~Grzechnik, K.~Syassen, I.~Loa, M.~Hanfland, J.~Y. Gesland, Phys. Rev. B 65 (2002) 104102.

\bibitem{Grzechnik05b}
A.~Grzechnik, K.~Friese, V.~Dmitriev, H.-P. Weber, J.-Y. Gesland, W.~A. Crichton, J. Phys.: Condens. Matter 17 (2005) 763--770.

\end{thebibliography}

\end{document}